\documentstyle[12pt]{article}
\title{NONLINEAR MAGNETOSONIC WAVES IN A MULTI-ION-SPECIES
PLASMA}
\author{S.~Boldyrev\\
{}~\\
{\em Princeton Plasma Physics Laboratory}\\ {\em P.O.Box 451, 
Princeton NJ 08543}}
\date{}
\begin{document}

\maketitle

\begin{abstract}

Magnetosonic waves are intensively studied due to their importance in space 
plasmas 
and also in fusion plasmas where they are used in particle acceleration and 
heating
experiments. In the present paper we investigate the magnetosonic waves 
propagating 
in a multi-ion-species plasma perpendicular to an external magnetic field.
Due to the presence of several ion species, this mode splits into two 
branches: high- and low-frequency modes. This opens a new channel of 
nonlinear 
interactions (between these two modes), and qualitatively changes the 
picture of 
turbulence in the long-wave region. Using the
exact kinetic approach, we derive a general system describing the 
propagation of 
nonlinearly coupled high- and low-frequency waves. This system includes
the~KdV, Boussinesq, and Zakharov equations as limiting cases. 
Solitary solutions of the system of coupled equations are obtained.

\end{abstract}

\newpage

\begin{centerline}
{I. DISPERSION RELATIONS FOR HIGH-FREQUENCY}
\end{centerline}
\begin{centerline}
{AND LOW-FREQUENCY WAVES}
\end{centerline}
\vskip1cm
We consider two-dimensional turbulence of magnetosonic (also referred to 
as
magnetohydrodynamic or compressional Alfven) waves, propagating 
perpendicular to an external magnetic field in a collisionless cold plasma 
with two ion species. As was shown in [1], in such a plasma the 
magnetosonic 
mode is split into two branches: low-frequency and high-frequency waves. 
The 
dispersion relation for them can be found from: 

\begin{eqnarray}
k^2=\frac{\omega^2}{c^2}\frac{\varepsilon^2_{\perp}-g^2}
{\varepsilon_{\perp}}\,\,,
\end{eqnarray}
\noindent where

\begin{eqnarray}
\varepsilon_{\perp}(\omega)=1-\Sigma_\alpha 
\frac{\omega^2_{p\alpha}}{\omega^2-\Omega^2_{\alpha}} \,\,\,\,\, , 
\,\,\,\,\,
g(\omega)= 
-\Sigma_{\alpha}\frac{\omega^2_{p\alpha}\Omega_\alpha}
{\omega(\omega^2-\Omega^2_{\alpha})} \,\,\,\,  ,
\end{eqnarray}

\noindent and $\alpha$  marks the particle type:  $\alpha =\lbrace a,b,e 
\rbrace $  (we denote ion species as $a$ and $b$). This equation has two 
solutions, one of which corresponds to a 
high frequency wave and the other to a low frequency wave [1]. The 
dispersion law of the high frequency wave is:

\begin{eqnarray}
\omega_{+} =\omega_0 + \mu_+k^2 \,\,\,,
\end{eqnarray}

\noindent where

\begin{eqnarray}
\omega_0 = \left (\frac{\omega^2_{pa}}{\Omega^2_a} 
+\frac{\omega^2_{pb}}{\Omega^2_b}\right )\Omega_a\Omega_b\Omega_e/ 
\omega^2_{pe} \,\,\,,
\end{eqnarray}

\begin{eqnarray}
\mu_+ = \frac{1}{2}\frac{\omega^2_{pa}\omega^2_{pb}}{\omega_0}\left ( 
\frac{1}{\Omega_a}-\frac{1}{\Omega_b}\right )^2 
\frac{\Omega^2_e}{\omega^4_{pe}}V_A^2 \,\,\,,
\end{eqnarray}

\noindent and $V_A$ is a modified Alfven velocity:

\begin{eqnarray} 
V_A^2= \frac{V_{Aa}^2V_{Ab}^2}{V_{Aa}^2+V_{Ab}^2}\,\,\,.
\end{eqnarray}
\noindent For the low frequency wave  Eq. (1) gives:
\begin{eqnarray}
\omega_- =V_Ak- \mu_- k^3 \,\,\,,
\end{eqnarray} 

\noindent where

\begin{eqnarray}
\mu_- = \frac{1}{2}V_A^3 \left (\frac{\omega^2_{pa}}{\Omega^2_a} 
+\frac{\omega^2_{pb}}{\Omega^2_b}\right )^{-2} \left [ 
\frac{\omega^2_{pa}\omega^2_{pb}}{\Omega^2_a\Omega^2_b}\left ( 
\frac{1}{\Omega_a}- \frac{1}{\Omega_b}\right )^2 \right ]\,\,\,.
\end{eqnarray}
\noindent These relations were established in [1]. 
Dispersion relations (3), (7) are obtained in the long wave region and 
are valid when $k<<max_{\alpha=a,b}\lbrace\omega_{p\alpha}/c\rbrace$. 
We will also need the polarization vectors for 
each of these modes. In a system where the wave vector is directed 
along the $x$-axis ($k_y=0$) we obtain:

\begin{eqnarray}
{\bf e}_\pm(k)= \frac{1}{({\tilde 
g}_\pm^2 +\tilde{\varepsilon}^2_{\perp\pm})^{1/2}} ( i\tilde g_\pm , 
-\tilde{\varepsilon}_{\perp\pm})\,\,\,,
\end{eqnarray}
\noindent where

\begin{eqnarray}
{\tilde g}_\pm (k)=g(\omega_\pm(k)) \,\,\,,
\end{eqnarray}

\begin{eqnarray}
{\tilde {\varepsilon}}_{\perp \pm}(k) = 
\varepsilon_{\perp}(\omega_{\pm}(k))\,\,\, . 
\end{eqnarray}
\vskip1cm

\begin{centerline}
{II. NONLINEAR CURRENT AND NONLINEAR RESPONSE TENSOR}
\end{centerline}

\vskip1cm

\noindent The nonlinear response $S_{kk_1k_2}$ is defined as:
\begin{eqnarray}
S_{kk_1k_2}= {\bf e}^*_i(k)S_{ijl}{\bf e}_j(k_1){\bf e}_l(k_2)\,\,\,,
\end{eqnarray}

\noindent and the tensor $S_{ijl}$ can be found from the nonlinear 
current:
\begin{eqnarray}
&j_{k,i}^N =  \int S_{ijl}(kk_1k_2)E_{k_1,j}E_{k_2,l} \delta({\bf k}-
{\bf k}_1-{\bf k}_2) \nonumber \\
& \times \delta (\omega - \omega_1 - 
\omega_2)\,d^2k_1\,d^2k_2\,d\omega_1\,d\omega_2 \,\, .
\end{eqnarray}

\noindent In our limit of cold plasma and small $k$ this current can 
be 
obtained from the collisionless kinetic equation, in which we neglect 
small terms $kV_{T\alpha}/\omega$ (where $V_{T\alpha}$ is the thermal 
velocity of particles of type $\alpha$). A simple iteration 
procedure~[4] 
then 
leads to the following expression for the nonlinear response tensor:
\begin{eqnarray}
S_{1ij}=\Sigma_\alpha \frac{-q_{\alpha}(\omega 
\delta_{i1}+i\Omega_{\alpha}\delta_{i2})} {4\pi 
m_\alpha(\Omega^2_{\alpha}-\omega^2)}k_{2s} 
(\varepsilon^{(\alpha)}_{sj}(k_2)-\delta_{sj}) \,\,, 
\end{eqnarray}

\begin{eqnarray}
S_{2ij}=\Sigma_\alpha \frac{-q_{\alpha}(\omega 
\delta_{i2}-i\Omega_{\alpha}\delta_{i1})} {4\pi 
m_\alpha(\Omega^2_{\alpha}-\omega^2)}k_{2s} 
(\varepsilon^{(\alpha)}_{sj}(k_2)-\delta_{sj}) \,\,. 
\end{eqnarray}
\noindent Particles of all types contribute to this tensor. As an 
example, let
us calculate the nonlinear response for the high-frequency field $E_1$ 
and
low-frequency field $E_2$. Consider the expression:

$$ k_{2s} \left [ \varepsilon^{(\alpha)}_{sj}(k)-\delta_{sj} \right ] 
e_j(k_2) $$

\noindent in our system where $k_{2y}=0$. We will always assume that 
the field with subscript $2$ corresponds to the low-frequency wave. 
Since 
for $\omega_2 = k_2V_A$ we 
have $\tilde g << {\tilde \varepsilon}_{\perp}$, we obtain that ${\bf 
e}_j(k_2) =(0,-1)$, and:

\begin{eqnarray}
k_{2s}\varepsilon^{(\alpha)}_{sj}e_j(k_2)= -ig^{(\alpha)}k_2 = -i 
\frac{\omega^2_{p\alpha}}{\Omega_\alpha V_A}\,\,. 
\end{eqnarray}

\noindent Further, for high-frequency waves ($\omega=\omega_0$) we 
have  ${\tilde \varepsilon}_{\perp} = \tilde g $ 
and therefore:

\begin{eqnarray}
{\bf e}^*(k)=-\frac{1}{\sqrt 2}e^{i\phi}(i,1),\,\,\,\,\, {\bf e}(k_1) 
= \frac{1}{\sqrt 2}e^{-i\phi_1} (i,-1)\,\,\, ,
\end{eqnarray}

\noindent where $\phi$ is the angle between the wave vector ${\bf k}$ 
and $x$-axis.

\noindent Now from (12), (14), (15) we get (with $B_0$ being the 
external magnetic field):

\begin{eqnarray}
S_{kk_1k_2} =  
-\frac{ic\omega_0}{4\pi B_0V_A}e^{i(\phi-\phi_1)}\,\,\,.
\end{eqnarray}
 
\noindent Note that in a general case we should use 
an expression symmetrized with respect to the last two 
indices for the nonlinear response: 
$$ {\tilde S}_{kk_1k_2}=S_{kk_1k_2} + S_{kk_2k_1} \,\,.$$
\noindent Calculation of $S_{kk_2k_1}$ can be done in an analogous 
way:

\begin{eqnarray}
S_{kk_2k_1} = -\frac{ic\omega_0}{4\pi B_0 
V_A} e^{i(\phi-\phi_1)} \beta k_1 \,\,\,,
\end{eqnarray}

\noindent where

$$ \beta =\Sigma_{\alpha}\frac{\omega^2_{p \alpha} 
\Omega_{\alpha}V_A}{2 \omega^2_0 (\omega_0+\Omega_{\alpha})^2} 
\,\,\,.$$

\noindent We will suppose that $\beta k_1 << 1$  and term 
(19) is 
negligible in comparison with (18). We will return to the meaning of 
this inequality in Section III.

\vskip1cm
\begin{centerline}
{III. KINETIC DERIVATON OF THE NONLINEAR COUPLED EQUATIONS}
\end{centerline}

\vskip1cm        
           
We use the kinetic approach for the derivation of  nonlinear
coupled equations for the high- and low-frequency waves. We use 
Maxwell's 
equations for the fields and the collisionless Boltzmann 
equation for the particle distribution function.

We represent the high-frequency field in the form
\begin{eqnarray}             
{\bf E}={\bf E^+}+{\bf E^-}\,\,\,,         
\end{eqnarray}             
               
\noindent where the field ${\bf E^+}$ is concentrated at the 
frequency 
$ \omega = \omega_0 $, and the field ${\bf E}^-$ at the 
frequency 
$ \omega = - \omega_0 $. Maxwell's equations give:    
             
\begin{eqnarray}          
\left( -k_ik_j + k^2 \delta_{ij} - \frac{ \omega^2}{c^2} 
\varepsilon_{ij}  
\right) E_j = \frac{i \omega 4 \pi }{c^2} j_i^N       
\end{eqnarray}            
              
\noindent where $j^N$ is the nonlinear current obtained from (13). 
This current is of 
the second order in field amplitudes, therefore if the field $E_j$ 
in equation 
(21) is concentrated at $ \omega_0 $, the only term contributing 
to the current is:    
$ \sim E^0 E^+ $, where $E^0$ is the  low-frequency field. In 
nonlinear current    
(21), one can also keep the higher order terms $ \sim E^+E^+E^- $. 
If the amplitude of 
the low-frequency field is small, $ E^0 \sim E^+E^- $, one should 
keep the cubic term
together with the quadratic one in (21). 
If the field $E_j$ in (13) 
is a low-frequency one, the main contribution to the nonlinear 
current is given 
by $E^0E^0$ and $E^+E^-$. 

Let us substitute  $E_j = e_j(k)E_{ \omega , k}$, where $ e_j(k)$    
is the the polarization vector of the field in (21), and multiply 
this expression 
by $e_i^*({\bf k})$. Then we have for the high-frequency field:

\begin{eqnarray}               
e_i^*({\bf k}) \left( - k_ik_j + k^2 \delta_{ij} - \frac{ 
\omega^2}{c^2}      
\varepsilon_{ij} \right) e_j(k) \equiv - \frac{ \omega^2}{c^2}        
( \varepsilon_{ \perp} -g)+ \frac{k^2}{2}\,\,\,,          
\end{eqnarray}                
                  
\noindent and, expanding  $ ( \varepsilon_{ \perp} -g) $ in small 
deviation of the frequency from $\omega_0$:           
               
\begin{eqnarray}             
\varepsilon_{ \perp} -g \simeq 0+( \omega - \omega_0)
\frac{ \partial }    
{ \partial \omega } ( \varepsilon_{ \perp } -g) =       
\frac{c^2}{2 \mu_+ \omega_0^2}( \omega - \omega_0) \,\,\,,      
\end{eqnarray}              
                
\noindent we get:       
                
\begin{eqnarray}              
( \omega - \omega_0 - \mu_+ k^2)E^+_{ \omega , k} =         
- \frac{8 \mu_+ \pi i \omega_0}{c^2} e_i^*(k)j^N_i(k)\,\,.        
\end{eqnarray}               
                  
\noindent This is the main equation for the high-frequency field. 
To close the 
equation, we express the nonlinear current in terms of the field 
amplitudes:        
                  
\begin{eqnarray}               
&j^N_i(k)=2 \int {\tilde S}_{ijk} E^+_J(k_1)E^0_k(k_2)\, 
d12 \nonumber \\      
&+6\int {\tilde \Sigma }_{ijkl} E^+_j(k_1)E^+_k(k_2)E^-_l(k_3) \,
d123 \,\,,      
\end{eqnarray}                 
                   
\noindent where we use the short-hand notation:        
               
$$               
d12 \equiv \delta ( \omega - \omega_1 - \omega_2)       
\delta ( {\bf k}-{\bf k}_1-{\bf k}_2)d \omega_1 d 
\omega_2 d{\bf k}_1    
d{\bf k}_2 \,\,\,,            
$$                
                
$$                
d123 \equiv \delta ( \omega - \omega_1 - \omega_2 - \omega_3)      
\delta ( {\bf k}-{\bf k}_1-{\bf k}_2 - {\bf k}_3)        
d \omega_1 d \omega_2 d \omega_3 d{\bf k}_1 d{\bf k}_2 d{\bf k}_3 
\,\,\,.     
$$                 
                 
\noindent The response $ {\tilde S}_{ijk}$ is symmetrized with 
respect to the last two indices,    
the the response $ {\tilde \Sigma }_{ijkl} $ with respect to the 
last three indices. 

The second order response can be obtained using the general 
formula (13), 
which gives:

\begin{eqnarray}            
{\tilde S}(kk_1k_2)=- \frac{ic \omega_0 k_2}{8 \pi B_0 \omega_2}   
e^{i(\phi - \phi_1)}(1+ O( \beta k_1))\,\,\,.       
\end{eqnarray}            
               
\noindent We will consider the limit $ \beta k_1 << 1$ (i.e.        
$k_1V_A << \frac{V_A^2}{c^2} \Omega_{ \alpha }$) and will neglect 
the last term in (26). One can also show that the third order 
term can be safely neglected for our problem. 

Consider now the derivation of the equation for the low-frequency 
field. Multiplying equation (21) by the complex conjugate vector 
of the low-frequency field polarization, we get:
              
\begin{eqnarray}           
\left( k^2 - \frac{\omega^2}{c^2} \varepsilon_{\perp } \right) 
E^0 =  \frac{4 \pi i \omega }{c^2}e^{*(0)}_i j_i^N \,\,\,.      
\end{eqnarray}            
              
\noindent This is the main equation for the low-frequency field. 
The contribution to the low-frequency current is given by two 
terms  
proportional to $E^0E^0$ and to $E^+E^-$. Consider the 
contribution 
of the first one. In this case, all three polarization vectors in  
formula (12) are related to the low-frequency field. Using 
result (16) and the 
following expression for the polarization vector ${\bf e}
({\bf k}_1)=
(\sin {\phi_1}, -\cos {\phi_1})$, we get:

\begin{eqnarray}              
S_1^{00} = \Sigma_{\alpha } - \frac{q_{\alpha }(\omega \sin 
{\phi_1}- i\Omega_{\alpha } \cos {\phi_1})}         
{4 \pi m_{\alpha }(\Omega_{\alpha }^2-\omega^2)}(-i)      
\frac{\omega_{p \alpha }^2k_2}{\Omega_{\alpha } \omega_2}
\,\,\,,   
\end{eqnarray}            
              
\begin{eqnarray}            
S_2^{00} = \Sigma_{\alpha } - \frac{q_{\alpha }(-\omega \cos 
{\phi_1}-   
i\Omega_{\alpha } \sin {\phi_1})}         
{4 \pi m_{\alpha }(\Omega_{\alpha }^2-\omega^2)}(-i)      
\frac{\omega_{p\alpha }^2k_2}{\Omega_{\alpha } \omega_2}\,\,\,.     
\end{eqnarray}             
                
\noindent Multiplying by           
${\bf e}^*({\bf k})=(\sin {\phi }, -\cos {\phi })$, we get:

\begin{eqnarray}          
S^{00} =\Sigma_{\alpha } \frac{i q_{\alpha } \omega_{p \alpha }^2k_2}   
{4 \pi m_{\alpha } \Omega_{\alpha } \omega_2 
(\Omega_{\alpha }^2 -\omega^2)}  
\lbrack \omega \cos {(\phi -\phi_1)}+i\Omega_{\alpha} 
\sin {(\phi_1- \phi )}  
\rbrack \,\,\,.            
\end{eqnarray}             
               
\noindent For $ \omega << \Omega_{\alpha}$, the second term in the 
brackets is 
small due to plasma quasineutrality, and we finally get: 

\begin{eqnarray}              
S^{00} \simeq \Sigma_{\alpha } \frac{i q_{\alpha } 
\omega_{p \alpha }^2k_2 \omega }   
{4 \pi m_{\alpha } \Omega_{\alpha }^3 \omega_2 }\cos {(\phi -\phi_1)}
\,\,\,.    
\end{eqnarray}              
                 
\noindent Now let us find the high-frequency field contribution 
to the 
low-frequency nonlinear current. We choose the system where 
${\bf k_2} =(k_2,\,\,0)$ and get: 
                
\begin{eqnarray}               
&{\bf e}^*({\bf k}) = (\sin {\phi }, -\cos {\phi }) \,\,\,, \\         
&{\bf e}^+({\bf k}_1) = \frac{1}{\sqrt {2}}e^{-i \phi_1}(i, -1)
\,\,\,, \nonumber \\    
&{\bf e}^-({\bf k}_2) = \frac{1}{\sqrt {2}}( -i, -1)\,\,\,.        
\end{eqnarray}                
                
\noindent The nonlinear response takes the form:         
                
$$               
S_1^{+-}(k_1k_2) = \Sigma_{\alpha } \frac{q_{\alpha }k_2}      
{8 \pi m_{\alpha } \Omega_{\alpha }}          
\frac{-\omega^2_{p \alpha}}{\omega_0 
(\Omega_{\alpha }+ \omega_0)}      
e^{-i \phi_1}\,\,\,,             
$$                 
                 
$$                 
S_1^{-+}(k_2k_1) = \Sigma_{\alpha } \frac{q_{\alpha }k_1}        
{8 \pi m_{\alpha } \Omega_{\alpha }}            
\frac{\omega^2_{p \alpha}}{\omega_0 (\Omega_{\alpha }+ \omega_0)}       
e^{i \phi_2}\,\,\,,              
$$                 
                  
$$             
S_2^{+-}(k_1k_2) = iS_1^{+-}(k_1k_2) \,\,\,,      
$$             
               
$$
S_2^{-+}(k_2k_1) = -iS_1^{-+}(k_2k_1) \,\,\,,      
$$             

\noindent and, multiplying by the polarization vector of the 
low-frequency field, we finally obtain: 

\begin{eqnarray}           
S^{0+-}(kk_1k_2)= - \frac{ic}{8 \pi B_0}        
(k_1 e^{i(\phi_2 - \phi )} + k_2 e^{i(\phi - \phi_1)}) \,\,\,.    
\end{eqnarray}             
               
\noindent We are now ready to write the system of coupled 
nonlinear equations. 
We will consider the one-dimensional case. In this case we 
will use the following
representation for the fields: ${\bf E}^0(\omega, k)=(0, -1)
E^0(\omega, k)$, ${\bf
E}^+(\omega, k)=\frac{1}{\sqrt{2}}(i, -1)E^+(\omega, k)$, ${\bf
E}^-(\omega, k)=\frac{1}{\sqrt{2}}(-i, -1)E^-(\omega, k)$. 
All of the
polarization vectors are chosen to be fixed, i.e. independent 
of the sign of $k$. 
Introduce the envelope of the high-frequency 
field ${\tilde E}^+$ as follows: $E^+(\omega, k) = 
{\tilde E}^+(\omega, k) 
e^{-i\omega_0 t}$. 
Then, in Fourier representation this system takes the form:        
                
\begin{eqnarray}             
&(\omega - \mu_+ k^2) {\tilde E}^+_{\omega , k} = - \frac{2 \mu_+ 
\omega^2_0}{c B_0}  
\int \frac{k_2}{\omega_2} {\tilde E}^+_1 E_2^0 d12 \nonumber \\     
&+ (third \,\,\, order \,\,\, terms) \,\,\,,        
\end{eqnarray}              
                 
\begin{eqnarray}              
&\left( k^2 - \frac{\omega^2}            
{c^2}\varepsilon_{\perp } \right)E^0_{\omega , k}=         
- \frac{ \omega k}{2 B_0 c}\int {\tilde E}^-_1 {\tilde E}^+_2 d12 
\nonumber \\     
& - \frac{1}{2} \frac{c \omega^2}{B_0 V_A^2}          
\int \lbrack \frac{k_1}{\omega_1} + \frac{k_2}{\omega_2}\rbrack 
E_1^0 E_2^0     
d12 \,\,\,.                
\end{eqnarray}                
                  
\noindent We need to introduce some ordering to understand what terms
 can be 
neglected in (35-36). First of all, the most interesting case is 
when the 
dispersion and the nonlinear terms in (35) are of the same order. 
This gives:

\begin{eqnarray}            
k^2 \sim \frac{\omega_0^2}{c B_0 V_0}E^0 \,\,\,,       
\end{eqnarray}            
               
\noindent where $V_0$ is the characteristic velocity of the 
nonlinear wave.
We also neglect the cubic term in (35). One can show that this 
can be justified only when

\begin{eqnarray}              
\frac{k^2c^2 V_0}{V_A^2 B_0} \vert E^+ \vert^2 <<        
\frac{\omega_0^2}{c}E^0 \,\,\,.            
\end{eqnarray}               
                  
\noindent Then in the left hand side of (36) we neglect the 
dispersion 
term $\frac{\mu_- k^4}{V_A} $, where          
$ \mu_- \sim \frac{V_A c^2}{\omega^2_{pa}} $. This is justified, 
since this term is 
always small compared to the last nonlinear term in (36) due to (37). 
Thus we 
assume (37) and (38). If one then introduces, for convenience, the 
magnetic field 
perturbation 
$ B_1 $ as $ E^0 = B_1 \frac{V_A}{c} $, two self-consistent 
orderings are possible:            
            
1. The first nonlinear term in (36) dominates the second one, i.e.     
$ B_1 << E^+ $;             
                
2. Both nonlinear terms in (36) are of the same order, i.e. 
$ B_1 \sim E^+ $.  

\noindent Consider the first ordering. From (37), (38)     
we get:               
                
\begin{eqnarray}              
B_0 \frac{V_A}{c} >> E^+ >> B_1 \,\,.          
\end{eqnarray}               
                 
\noindent Then, the linear terms in (36) must be of the same order 
as the nonlinear 
terms. Comparing $k^2 E^0$ and the nonlinear term, we have:

\begin{eqnarray}            
B_1 \sim \frac{V_0}{V_A} \left( \frac{E^+}{B_0} \right) E^+ \,\,\,,   
\end{eqnarray}            
              
\noindent which is in agreement with (39). 
Consider the left half of inequality (39):         
               
\begin{eqnarray}            
 \frac{E^+}{B_0} << \frac{V_A}{c} \,\,\,,        
\end{eqnarray}             
                
\noindent Using~(40), it is easy to check that this condition 
coincides with the 
previous assumption $\beta k << 1$ (see~(49)). Thus, in limit~(39), 
we neglect the 
cubic term in equation~(35) and the second nonlinear term in 
equation~(36).

Consider now the second ordering, $B_1 \sim E^+$.
In this case we keep both nonlinear terms in equation (36). 
Comparing 
these nonlinear terms with the linear ones, we find that they are 
of the same 
order if the velocity of the nonlinear wave $V_0$ is close 
to $V_A$:

$$                  
\frac{\delta V}{V_A} \sim \frac{E^+}{B_0} \,\,\,, \,\,\,        
\delta V = \vert V_A - V_0 \vert \,\,\,.            
$$                  
                   
\noindent The condition $\beta k << 1$ together with  (37) then 
gives:   
             
\begin{eqnarray}           
\frac{E^+}{B_0} << \frac{V_A^2}{c^2} \,\,\,,       
\end{eqnarray}            
              
\noindent which is more restrictive than (41). We therefore 
demand (42), 
and the second ordering is then also self-consistent. 
In this case, we neglect the cubic term in (35) and keep both 
nonlinear 
terms in (36).

\vskip1cm
\begin{centerline}
{IV. SOLITONS IN A MAGNETIZED MULTI-ION PLASMA}
\end{centerline}
\vskip1cm

Consider the soliton solutions for system (35-36). Let us rewrite 
the system in the $(x,t)$-representation. For this purpose we 
introduce 
the ``potential'' $\Phi$: $E^0 = \partial \Phi / \partial t$. The 
system 
then takes the form:

\begin{eqnarray}            
\left( i \frac{\partial }{\partial t} + \mu_+ \frac{\partial^2}
{\partial x^2}  
\right){\tilde E}^+ = \frac{2 \mu_+ \omega^2_0}{cB_0} {\tilde E}^+    
\frac{\partial \Phi }{\partial x} \,\,\,,        
\end{eqnarray}             
               
\begin{eqnarray}             
\left( \frac{1}{V_A^2} \frac{\partial^2}{\partial t^2} -      
\frac{\partial^2}{\partial x^2} \right) \Phi = - \frac{1}{2B_0 c}     
\frac{\partial }{\partial x} \vert {\tilde E}^+ \vert^2 -       
\frac{c}{B_0 V_A^2} \frac{\partial }{\partial t}(\Phi_x \Phi_t) 
\,\,\,.    
\end{eqnarray}            
              
\noindent As we have already seen, two limiting cases are possible, 
depending on the 
relative magnitudes of the nonlinear terms in (44). In the first case 
we neglect the 
second nonlinear term in (44) and get the following system:

\begin{eqnarray}            
\left( i \frac{\partial }{\partial t} + \mu_+ \frac{\partial^2}
{\partial x^2}  
\right){\tilde E}^+ = \frac{2 \mu_+ \omega^2_0}{cB_0} {\tilde E}^+    
\frac{\partial \Phi}{\partial x} \,\,\,,         
\end{eqnarray}             
                
\begin{eqnarray}             
\left( \frac{1}{V_A^2} \frac{\partial^2}{\partial t^2} -       
\frac{\partial^2}{\partial x^2} \right) \Phi = - \frac{1}{2B_0 c}   
\frac{\partial }{\partial x} \vert {\tilde E}^+ \vert^2 \,\,\,.   
\end{eqnarray}            

\noindent In the second case we keep both nonlinear terms in (44). 
In this case, 
the nonlinear wave has velocity $V_0$ which is  close 
to the velocity $V_A$, and we can replace: $\frac{\partial^2}
{\partial t^2} =    
V_A^2 \frac{\partial^2}{\partial x^2}$ in the right hand side of 
equation (44). 
Then, if we are interested in the wave moving in some particular 
direction (say, to the right) we can further simplify (44) by writing:

$$  
 \frac{1}{V_A^2} \frac{\partial^2}{\partial t^2} -       
\frac{\partial^2}{\partial x^2}=           
\left( \frac{1}{V_A}\frac{\partial }          
{\partial t} - \frac{\partial }{\partial x} \right)       
\left( \frac{1}{V_A}\frac{\partial }          
{\partial t} + \frac{\partial }{\partial x} \right)    
 \simeq -2 \frac{\partial }{\partial x}          
\left( \frac{1}{V_A}\frac{\partial }           
{\partial t} + \frac{\partial }{\partial x} \right) \,\,\,.   
$$                

\noindent We then  get the following system:          
                  
\begin{eqnarray}               
\left( i \frac{\partial }{\partial t} + \mu_+ \frac{\partial^2}
{\partial x^2}     
\right){\tilde E}^+ = \frac{2 \mu_+ \omega^2_0}{cB_0} {\tilde E}^+   
\frac{\partial \Phi }{\partial x} \,\,\,,        
\end{eqnarray}            
               
\begin{eqnarray}            
\left( \frac{1}{V_A}\frac{\partial }         
{\partial t} + \frac{\partial }{\partial x} \right) \Phi_x =     
\frac{1}{4 B_0 c} \frac{\partial }{\partial x} \vert {\tilde E}^+ 
\vert^2   
+ \frac{c}{2 B_0}(\Phi_x^2)_x \,\,\,.          
\end{eqnarray}              

Let us look for the stationary solutions of systems (45-46) and 
(47-48).
It is convenient to represent the fields in the form:

\begin{eqnarray}            
&{\tilde E}^+ = E(\theta ) \exp ( \frac{1}{2\mu_+}i V_0 x -     
\frac{1}{4 \mu_+}i V_0^2 t - i \mu_+ \Delta t ) \,\,\,, \nonumber \\    
& \Phi = \Phi (\theta )\,\,\,,          
\end{eqnarray}             
               
\noindent where $\theta = x - V_0 t$, and $V_0 , \,\,\, \Delta $ 
-- some parameters.   
First consider system (45-46). Substituting expression (49) into it, 
we find:
                
\begin{eqnarray}              
E \Delta + E_{\theta \theta}           
= \frac{2 \omega^2_0}{c B_0} E \Phi_{\theta}\,\,\,,        
\end{eqnarray}              
                 
\begin{eqnarray}               
\left( \frac{V_0^2}{V_A^2} - 1 \right) \Phi_{\theta \theta } =       
- \frac{1}{2 B_0 c}(E^2)_{\theta } \,\,\,.           
\end{eqnarray}               
                  
\noindent Integrating equation (51) once, we get:         
                  
\begin{eqnarray}                
\Phi_{\theta } = \frac{ V_A^2}{2 B_0 c (V_a^2 - V_0^2) }E^2 + A 
\,\,\,,  
\end{eqnarray}           
              
\noindent where $A$ is some constant of integration. 
Substituting this result into (50), we consider two cases: 
$V_0 < V_A$ and $V_0 > V_A$. In the first case, the soliton solution 
has the form: 

\begin{eqnarray}             
E = \sqrt {\frac{B}{C}} th (\sqrt {\frac{B}{2}} \theta ) \,\,\,,    
\end{eqnarray}             
               
\noindent where $B \equiv \Delta - \frac{2 \omega_0^2}{c B_0} A$,     
$ C \equiv \frac{ \omega_0^2 V_A^2}{B_0^2 c^2 (V_A^2 - V_0^2)}$. 
The soliton 
solution exists only when  $B > 0$. In the opposite limit 
$V_0 > V_A$, we 
have another soliton solution.

\begin{eqnarray} 
E = \sqrt {\frac{2B}{C}} \frac{1}{ch(\sqrt {\vert B \vert } \theta )} 
\,\,\,, 
\end{eqnarray} 

\noindent which exists only for $B < 0$. 

Note, that system (45-46) is analogous to the Zakharov system for 
coupled 
Langmuir and ion-acoustic waves [3]. The difference is in the sign 
of the 
nonlinear 
term in equation (45). Consider now system (47-48). Substituting (49) 
into (48) we obtain:

\begin{eqnarray}                                   
\left( \frac{V_A - V_0}{V_A} \right) \Phi_{\theta } = 
\frac{1}{4 B_0 c}E^2       
+ \frac{c}{2 B_0} (\Phi_{\theta })^2 + F \,\,\,,                    
\end{eqnarray}                                      
                                              
\noindent where $F$ is an arbitrary constant of integration. The soliton 
solution takes the form:                                      
                                                
\begin{eqnarray}                                        
E = \frac{a th (b \theta )}{ch (b \theta )} \,\,\,, \nonumber \\                 
\Phi_{\theta } = \frac{d}{ch^2(b \theta )} + D \,\,\,,                      
\end{eqnarray}                                           
                                                  
\noindent where                                            
                                                   
$$                                                   
b^2 = -( \Delta - \frac{2 \omega_0^2}{cB_0} D )                             
\equiv - {\tilde {\Delta }} \,\,\,,                                   
$$                                                    
                                                      
$$                                                     
d = \frac{3cB_0}{\omega_0^2} {\tilde {\Delta }} \,\,\,,                           
$$                                                      
                                                        
$$ 
\frac{V_A -V_0}{V_0} = \frac{Dc}{B_0}         
+ \frac{3c^2}{2 \omega_0^2} {\tilde {\Delta }} \,\,\,,  
$$                            
                              
$$                             
a^2 = \frac{18 B_0^2 c^4 {\tilde {\Delta }}^2}{\omega_0^4} \,\,\,  
$$

\noindent This solution depends on two arbitrary parameters, 
$ \Delta $ and 
$D$, is valid for both $V_0 < V_A$ and $V_0 > V_A$, is 
analytical at 
$V_0 = V_A$, and exists only for $ \tilde {\Delta } < 0$. 

\vskip1cm
\begin{centerline}
{V. CONCLUSION}
\end{centerline}
\vskip1cm

In conclusion, we have presented the general kinetic method for 
derivation of
nonlinear equations for magnetosonic waves, propagating 
perpendicular to an external
magnetic field in a multi-ion plasma. Equations~(45)-(46) 
constitute the so-called 
Zakharov system [3], first derived for Langmuir and ion-acoustic 
waves 
in non-isotermal ($T_e \gg T_i$)
plasma. Without the high-frequency field and with dispersion 
term $\mu_- k^4/V_A$,
Eq.~(44) is the Boussinesq equation for acoustic waves, and it can 
be reduced to 
the KdV equation exactly in the same way as we went from (44) to (48).

Equations (45)-(46) can be used to describe weak turbulence of 
interacting high- and
low-frequency waves. For this purpose, the equations should be 
rewritten in the 
Hamiltonian
form, expanded in small amplitudes of interacting fields, and 
averaged over
statistical realizations. Such a procedure is described in [5]. 
In [2] the same turbulence was considered using another
method. The system can also be used to describe transition to 
strong turbulence.

The system analogous to (47)-(48) was considered by Makhankov [6], and
also by Nishikawa {\em et al}~[7] in connection with near-sonic solitons 
of coupled 
Langmuir and ion-acoustic waves.

We would like to stress that the developed method, though it is not 
as transparent as
the hydrodynamical approach, is exact and rather straightforward, 
and can be easily 
generalized to other plasma systems. Moreover, it, in principle, 
allows the 
consideration of such kinetic effects as Landau damping on a 
rigorous basis, which
is impossible with the hydrodynamical approach. 
 

\vskip1cm This work was supported by U.S.D.o.E Contract No. 
DE-AC02-76-CHO-3073.

\end{document}